\documentclass[pra,twocolumn,superscriptaddress,showpacs,a4paper]{revtex4}
\usepackage{graphicx}
\usepackage{graphics}
\usepackage{amssymb}
\usepackage{amsmath}
\usepackage{epsfig}
\usepackage{latexsym}
\usepackage{color}
\usepackage{rotating}

\begin{document}

\title{Entangling a nanomechanical resonator and a superconducting microwave cavity}
\author{D. Vitali}
\affiliation{Dipartmento di Fisica, Universit\`{a} di Camerino,
I-62032 Camerino (MC) Italy}
\author{ P. Tombesi}
\affiliation{Dipartmento di Fisica, Universit\`{a} di Camerino,
I-62032 Camerino (MC) Italy}
\author{M. J. Woolley}
\affiliation{Department of Physics, School of Physical Sciences, The
University of Queensland, St Lucia, QLD 4072, Australia}
\author{A. C. Doherty}
\affiliation{Department of Physics, School of Physical
Sciences, The University of Queensland, St Lucia, QLD 4072,
Australia}
\author{G. J. Milburn}
\affiliation{Department of Physics, School of Physical Sciences, The
University of Queensland, St Lucia, QLD 4072, Australia}

\date{\today}

\begin{abstract}
We propose a scheme able to entangle at the steady state a nanomechanical resonator with a microwave cavity mode of a driven superconducting
coplanar waveguide. The nanomechanical resonator is capacitively coupled with the central conductor of the waveguide and stationary entanglement
is achievable up to temperatures of tens of milliKelvin.
\end{abstract}


\maketitle

\section{Introduction}

Entanglement is one of a number of inherently quantum phenomena that it is hoped will soon be observable in macroscopic mechanical systems
\cite{Blencowe1}. Aside from the interest in studying quantum mechanics in a new regime, entanglement may be used as part of read-out schemes in
quantum information processing applications. Methods for entangling a nanomechanical resonator with a Cooper pair box \cite{armour1}, or an
optical mode \cite{Vitali07}, for entangling two charge qubits \cite{zou1} or two Josephson junctions \cite{cleland1} via nanomechanical
resonators, and for entangling two nanomechanical resonators via trapped ions \cite{tian1}, Cooper pair boxes \cite{tian2}, entanglement
swapping \cite{Pir06}, and sudden switching of electrical interactions \cite{plenio1}, have all been proposed. In the earliest proposal,
\cite{armour1}, the entanglement provided a means for measuring the decoherence rate of coherent superpositions of nanomechanical resonator
states. More recently, a scheme for entangling a superconducting coplanar waveguide field with a nanomechanical resonator, via a Cooper pair box
within the waveguide \cite{ringsmuth}, was proposed.

Here we propose a different scheme for entangling the nanomechanical resonator, based on the capacitive coupling of the resonator with the
central conductor of the superconducting, coplanar waveguide, and which does not require any Cooper pair box (see Ref.~\cite{tian3} for a
related proposal). The paper is organized as follows. In Sec. II we derive the Quantum Langevin equations (QLE) of the system and discuss when
they can be linearized around the semiclassical steady state. In Section III we study the steady state of the system and quantify its
entanglement by using the logarithmic negativity, while Section IV is for conclusions.

\section{Quantum Langevin equations and their linearization}

\begin{figure}[h]
\includegraphics[scale=0.5]{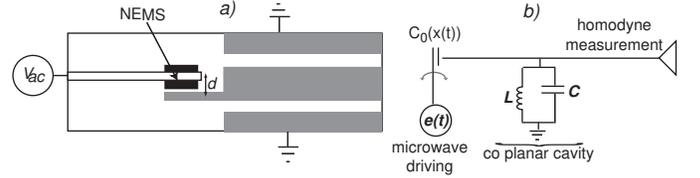}
\caption{Schematic representation of the capacitive coupling between a nanomechanical resonator and a superconducting coplanar microwave cavity.
(a) Plan view of the device; the dark region is etched away and the resulting cantilever is metalized to form one plate of a capacitor. (b)
Equivalent circuit.} \label{fig1}
\end{figure}

The proposed scheme is shown in Figure~\ref{fig1}: a nanomechanical resonator is capacitively coupled to the central conductor of a
superconducting, coplanar waveguide that forms a microwave cavity of resonant frequency $\omega_c$. The cavity is driven at a frequency
$\omega_0 = \omega_{c}-\Delta_0$. In view of the equivalent circuit, the effective Hamiltonian for the coupled system is
\begin{equation}
H=\frac{p_x^2}{2m}+\frac{m\omega_m^2 x^2 }{2}+\frac{\Phi^2}{2L}+\frac{Q^2}{2(C+C_0(x))}-e(t)Q,
\end{equation}
where $(x,p_x)$ are the canonical position and momentum of the resonator, and $(\Phi,Q)$ are the canonical coordinates for the cavity,
representing respectively the flux through an equivalent inductor $L$ and the charge on an equivalent capacitor $C$.
The coherent driving of the cavity is given by the electric potential $e(t)= -i\sqrt{2\hbar\omega_c L}E(e^{i\omega_0t} -e^{-i\omega_0t}) $.
The function $C_0(x)$ describes the capacitive coupling between the cavity and the resonator, as a function of the resonator displacement $x$.
Expanding this around the equilibrium position of the resonator at $d$ from the cavity and with capacitance $C_0$, we have $C_0(x)=C_0(1-x(t)/d)$.
Expanding the capacitive energy as a Taylor series, we find to first order,
\begin{equation}
\frac{Q^2}{2(C+C_0(x))} = \frac{Q^2}{2C_\Sigma} +  \frac{\beta}{2d C_\Sigma}x(t)Q^2,
\end{equation}
where $C_\Sigma= C+C_0$ and $\beta=C_0/C_\Sigma$.

We can now quantize the Hamiltonian, promoting the canonical coordinates to operators with $[\hat{x},\hat{p}_x]=[\hat{Q},\hat{\Phi}]=i\hbar$.
The quantum Hamiltonian, in terms of the raising and lowering operators for the cavity ($a^\dagger,a$) and the resonator dimensionless canonical
operators $(\hat{q},\hat{p})$, is
\begin{eqnarray}
H&=&\hbar\omega_c a^\dagger a+\frac{\hbar\omega_m}{2}(\hat{q}^2+\hat{p}^2)+\hbar \frac{G_0}{2}\hat{q}(a+a^\dagger)^2 \nonumber \\
&& \ \ -i\hbar E(e^{i\omega_0 t}-e^{-i\omega_0 t})(a+a^\dagger),
\end{eqnarray}
where
\begin{eqnarray}
a & = & \sqrt{\frac{\omega_c L}{2\hbar}} \hat{Q}+ \frac{i}{\sqrt{2\hbar \omega_c L}}\hat{\Phi},\\
\hat{q} & = & \sqrt{\frac{m\omega_m}{\hbar}} \hat{x}, \;\;\hat{p} = \frac{\hat{p}_x}{\sqrt{\hbar m\omega_m}},
\end{eqnarray}
and the coupling depends on
\begin{equation} \label{coupling}
G_0 = \beta\omega_c\left (\frac{1}{2d} \sqrt{\frac{\hbar}{m\omega_m}}\right ).
\end{equation}
Typically, $\omega_c \gg \omega_m$ since $\omega_c/2\pi\approx 10$ GHz \cite{schoelkopf}, while $\omega_m/2\pi \approx 20$ MHz \cite{schwab}. It
is convenient to move into an interaction picture with respect to $\hbar \omega_0 a^\dagger a$, and neglect terms oscillating at $\pm
2\omega_0$. The resulting Hamiltonian is
\begin{equation} \label{Hint}
H_I=\hbar\Delta_0 a^\dagger a+\frac{\hbar\omega_m}{2}(\hat{q}^2+\hat{p}^2)+\hbar G_0 \hat{q}a^\dagger a-i\hbar E(a-a^\dagger).
\end{equation}
The coupling term represents a low frequency modulation of the cavity resonance frequency. This will cause a phase modulation of the cavity
field and write sidebands onto the cavity spectrum at multiples of $\omega_m$ from $\omega_c$.

The resonator has a mechanical damping rate $\gamma_m$ and the cavity bandwidth is $2\kappa$. System dynamics also depend on the cavity
input noise $a^{in}(t)$, where
\begin{equation}
\langle a^{in,\dag}(t)a^{in}(t')\rangle =\bar{n}_a \delta (t-t'), \label{input2}
\end{equation}
with $\bar{n}_a=\left(\exp\{\hbar \omega_c/k_BT\}-1\right)^{-1}$, and also on the Brownian noise acting on the cavity ends $\xi (t)$,
with correlation function \cite{GIOV01}
\begin{equation}\label{browncorre}
\left \langle \xi(t) \xi(t')\right \rangle = \frac{\gamma_m}{\omega_m} \int \frac{d\omega}{2\pi} e^{-i\omega(t-t')} \omega
\left[\coth\left(\frac{\hbar \omega}{2k_BT}\right)+1\right].
\end{equation}
Clearly, $\xi(t)$ is not delta-correlated and does not describe a Markovian process. However, quantum effects are achievable only when using
resonators with a large mechanical quality factor $(\mathcal{Q}_m =\omega_m/\gamma_m \gg 1)$, and in this limit $\xi(t)$ becomes
delta-correlated~\cite{benguria},
\begin{equation}\label{browncorre6}
\left \langle \xi(t) \xi(t')+\xi(t') \xi(t)\right \rangle/2 \simeq \gamma_m \left(2\bar{n}_b+1\right) \delta(t-t'),
\end{equation}
where $\bar{n}_b=\left(\exp\{\hbar \omega_m/k_BT\}-1\right)^{-1}$, and we recover a Markovian process. Adding these inputs to the
equations of motion that follow from (\ref{Hint}), we obtain the nonlinear quantum Langevin equations (QLEs)
\begin{subequations}
\label{nonlinlang}
\begin{eqnarray}
 \dot{q}&=&\omega_m p, \\
 \dot{p}&=&-\omega_m q - \gamma_m p + G_0  a^{\dag}a +\xi, \\
 \dot{a}&=&-(\kappa+i\Delta_0)a +i G_0 a q +E+\sqrt{2\kappa} a^{in}.
\end{eqnarray}
\end{subequations}
Neglecting the noise and treating the deterministic equations as classical, with $a\rightarrow\alpha$ a complex field amplitude, we find the
fixed points of the system by setting the left hand side of Eqs.(\ref{nonlinlang}) to zero.  The fixed points are then given by
\begin{eqnarray}
\label{steadystate1}
p_s & = & 0 ,\\
\label{steadystate2}
q_s & = & \frac{G_0}{\omega_m}n_s,\\
\label{steadystate3} E^2 & = & n_s\left (\kappa^2+(\Delta_0-G_0^2n_s/\omega_m)^2\right ),
\end{eqnarray}
where the steady state photon number in the cavity is defined as  $n_s=|\alpha_s|^2$. Eq.~(\ref{steadystate3}) is the same as the equation of
state for optical bistability in a dispersive non linear medium \cite{WallsMilb} and thus we expect for $\Delta_0>0$ there will be multiple
stable fixed points.

The quantum dynamics of the full nonlinear system is difficult to analyze so we linearize around the semiclassical fixed points. That is, we
write $a=\alpha_s+\delta a$, $q=q_s+\delta q$ and $p=p_s+\delta p$. This decouples our system into a set of nonlinear algebraic equations for
the steady-state values and a set of QLEs for the fluctuation operators. The steady-state values are given by Eqs.~(\ref{steadystate1}) and
(\ref{steadystate2}), and $\alpha_s = E/(\kappa+i \Delta)$; an implicit equation for $\alpha_s$, since the effective detuning $\Delta$ is given
by $\Delta = \Delta_0- G_0^2 |\alpha_s|^2/\omega_m$. The QLEs for the fluctuations are
\begin{subequations}
\label{nlqles}
\begin{eqnarray}
\delta \dot{q}&=&\omega_m \delta p,\\
\delta \dot{p}&=&-\omega_m \delta q - \gamma_m \delta p + G_0 \left(\alpha_s \delta a^{\dag}+ \alpha_s ^* \delta a \right)\nonumber \\
&& + \delta a^{\dag}
\delta a + \xi, \label{nlqlesb}\\
\label{mode} \delta \dot{a}&=&-(\kappa+i\Delta)\delta a +i G_0 \left(\alpha_s + \delta a\right) \delta q +\sqrt{2\kappa} a^{in} \label{nlqlesc}.
\end{eqnarray}
\end{subequations}
Provided the cavity is driven intensely, $|\alpha_s| \gg 1$, we can safely neglect the terms $\delta a^{\dag} \delta a$ in Eq.~(\ref{nlqlesb})
and $ \delta a \delta q$ in Eq.~(\ref{nlqlesc}), and obtain the linearized QLEs
\begin{subequations}
\label{lle3}
\begin{eqnarray}
\delta \dot{q}&=&\omega_m \delta p, \label{lle3a} \\
\delta \dot{p}&=&-\omega_m \delta q - \gamma_m \delta p + G_0 \alpha_s \left( \delta a^{\dag}+ \delta a \right)+ \xi ,\label{lle3b} \\
\label{modelin} \delta \dot{a}&=&-(\kappa+i\Delta)\delta a +i G_0 \alpha_s  \delta q +\sqrt{2\kappa} a^{in},
\end{eqnarray}
\end{subequations}
where we have chosen the phase reference so that $\alpha_s$ can be taken as real.

\section{Steady state of the system and its entanglement properties}

In order to characterize the steady state of the system, it is convenient to rewrite Eqs.~(\ref{lle3}), defining $G=G_0 \alpha_s \sqrt{2}$, in
terms of the field quadratures $\delta X = (\delta a + \delta a^\dagger)/\sqrt{2}$ and $\delta Y = -i(\delta a - \delta a^\dagger)/\sqrt{2} $,
that is,
\begin{subequations}
\label{lle1}
\begin{eqnarray}
\delta \dot{q}&=&\omega_m \delta p ,\\
\delta \dot{p}&=&-\omega_m \delta q - \gamma_m \delta p + G \delta X +\xi, \\
\delta \dot{X}&=&-\kappa \delta X+\Delta \delta Y +\sqrt{2\kappa} X^{in},  \\
\delta \dot{Y}&=&-\kappa \delta Y-\Delta \delta X +G\delta q +\sqrt{2\kappa}  Y^{in},
\end{eqnarray}
\end{subequations}
where $X^{in}= (\delta a^{in} + \delta a^{in,\dagger})/\sqrt{2}$ and $ Y^{in} = -i(\delta a^{in} - \delta a^{in,\dagger})/\sqrt{2} $. In matrix
form, Eqs.~(\ref{lle1}) can be written as
\begin{equation} \label{compact}
\dot{u}(t)=A u(t)+n(t),
\end{equation}
where $u^{T}(t) =(\delta q(t), \delta p(t),\delta X(t), \delta Y(t))$, $n^{T}(t) =(0, \xi (t), \sqrt{2\kappa}X^{in}(t), \sqrt{2\kappa}Y^{in}(t))$ and
\begin{equation}\label{dynmatrwa}
  A=\left(\begin{array}{cccc}
    0 & \omega_m & 0 & 0 \\
     -\omega_m & -\gamma_m & G & 0 \\
    0 & 0 & -\kappa & \Delta \\
    G & 0 & -\Delta & -\kappa
  \end{array}\right),
\end{equation}
Eq.~(\ref{compact}) has the solution
\begin{equation} \label{dynsolrwa}
u(t)=M(t) u(0)+\int_0^t ds M(s) n(s),
\end{equation}
where $M(t)=\exp(A t)$. The stability conditions can be derived by applying the Routh-Hurwitz criterion \cite{grad},
\begin{subequations} \label{stab}
\begin{eqnarray}
s_1&=&2\gamma_m\kappa\left\{  \left[  \kappa^{2}+\left(\omega_m-\Delta\right) ^{2}\right] \left[  \kappa^{2}+\left(
\omega_m+\Delta\right) ^{2}\right] \right. \nonumber \label{stab1}\\
&&\left.+\gamma_m\left[  \left( \gamma_m+2\kappa\right)  \left( \kappa^{2}+\Delta ^{2}\right) +2\kappa\omega_{m}^{2}\right] \right\} \nonumber \\
&&+\Delta\omega_{m} G^{2}\left(  \gamma_m+2\kappa\right)  ^{2}>0, \\
s_2&=&\omega_m\left(\Delta^2+\kappa^2\right)- G^2 \Delta > 0.
\end{eqnarray}
\end{subequations}
For driving on the blue sideband of the cavity $(\Delta = -\omega_m)$ we have
\begin{eqnarray} \label{bluestab}
G & < & \left\{ 2\gamma_m\kappa \left(  \kappa^4 +4\omega^2_m\kappa^2+\gamma^2_m\kappa^2+4\gamma_m\kappa\omega^2_m  \right.\right. \nonumber \\
& & \left.\left. \ \ +2\gamma_m\kappa^3 +\gamma^2_m\omega^2_m \right)\right\}^{\frac{1}{2}} / \left\{ \omega_m (\gamma_m +2\kappa) \right\},
\end{eqnarray}
while for driving on the red sideband $(\Delta = \omega_m)$,
\begin{equation} \label{redstab}
G<\sqrt{\omega^2_m+\kappa^2}.
\end{equation}
Since the noise terms in Eq.~(\ref{compact}) are zero-mean Gaussian and the dynamics are linear, the steady-state for the fluctuations is a
two-mode Gaussian state, fully characterized by its symmetrically-ordered $4 \times 4 $ correlation matrix. This has components $ V_{ij}=\langle
u_i(\infty)u_j(\infty)+ u_j(\infty)u_i(\infty)\rangle/2$. When the system is stable, using (\ref{dynsolrwa}), we get
\begin{equation} \label{cm2rwa}
V_{ij}=\sum_{k,l}\int_0^{\infty} ds \int_0^{\infty}ds'M(s)_{ik}M(s')_{jl}\Phi(s-s')_{kl},
\end{equation}
where $\Phi(s-s')_{kl}=\langle n(s)_k n(s')_l+ n(s')_l n(s)_k\rangle/2$ is the matrix of stationary noise correlation functions. Here
$\Phi(s-s')_{kl}= D_{kl}\delta(s-s') $, where $D=\mathrm{Diag}[0,\gamma_m (2\bar{n}_b+1), 2\kappa(\bar{n}_a+1/2),2\kappa(\bar{n}_a+1/2)]$ and
(\ref{cm2rwa}) becomes
\begin{equation} \label{cm3rwa}
V =\int_0^{\infty} ds  M(s)DM(s)^{T}
\end{equation}
which, by Lyapunov's first theorem \cite{parks}, is equivalent to
\begin{equation} \label{lyaprwa}
AV+VA^{T}=-D.
\end{equation}
Solving this equation, we can then quantify the entanglement of the steady-state by means of the logarithmic negativity, $E_{\mathcal{N}}$
\cite{werner,Salerno1}. This entanglement measure is particularly convenient because it is the only one which can always be explicitly computed
and it is also additive \cite{supernote}. In the continuous variable case we have
\begin{equation}
E_{\mathcal{N}}=\max [0,-\ln 2\eta ]  \label{logneg}
\end{equation}
where $ \eta = 2^{-1/2}\left[ \Sigma (V)- \left[\Sigma (V)^{2}-4\det V\right] ^{1/2}\right]^{1/2}$, with $\Sigma (V)$ expressed in terms of the
$2\times2$ block matrix
\begin{equation}
 V=\left[\begin{array}{cc} V_{b} & V_{corr} \\ V_{corr}^{T} & V_{a} \end{array}\right] \label{blocks}
\end{equation}
as $\Sigma (V) = \det V_{b}+\det V_{a}-2\det V_{corr}$.

The logarithmic negativity, assuming $\Delta =\pm \omega_m$, depends on $T$, $G$, $\omega_c$, $\kappa$, $\omega_m$ and $\gamma_m$. We first
consider the zero-temperature entanglement, such that our results are independent of $T$ and $\omega_c$. In all cases, entanglement increases
with increasing coupling $G$; the limit on our entanglement being due to the limit on $G$ specified by our stability conditions,
(\ref{bluestab}) and (\ref{redstab}), and we shall set $G$ just below this threshold. At zero temperature, the absolute magnitude of $\omega_m$
is also insignificant, so we may hold it fixed (at $\omega_m = 5\times 10^8s^{-1}$, say), leaving $\kappa$ and $\gamma_m$ as our remaining free
parameters. It is implicitly assumed here that damping rates are controllable independent of resonant frequencies. The zero-temperature
$E_\mathcal{N}$ is shown in Figures 2(a) and 2(b) for driving on the blue and red sidebands, respectively. From this data, along with the
stability conditions, we note that on the blue sideband, entanglement is maximized in a regime where $\omega_m \gg G,\kappa$. This is not the
case for driving on the red sideband. We also observe that the logarithmic negativity plateaus in both cases as $\kappa$ and $\gamma_m$
increase.

Now the temperature dependence of the entanglement follows from the Planck distributions specifying the noise input correlation functions,
Eqs.~(\ref{input2}) and (\ref{browncorre6}); hence the magnitudes of the resonant frequencies become significant in these calculations. Typical
temperature dependence of the logarithmic negativity is shown in Figure 3(a), decreasing from a positive value at zero temperature to zero at a
temperature we shall refer to as the critical temperature, $T_c$. Now $T_c$ increases both with increasing $\omega_c$ and $\omega_m$;
henceforth, we shall consider these fixed, with $\omega_c = 10^{10}s^{-1}$. The dependence of $T_c$ on the damping is shown in Figures 3(b) and
3(c); the entanglement in the red sideband case appears more robust with respect to increases in temperature.

\begin{figure*}[th!]
\includegraphics{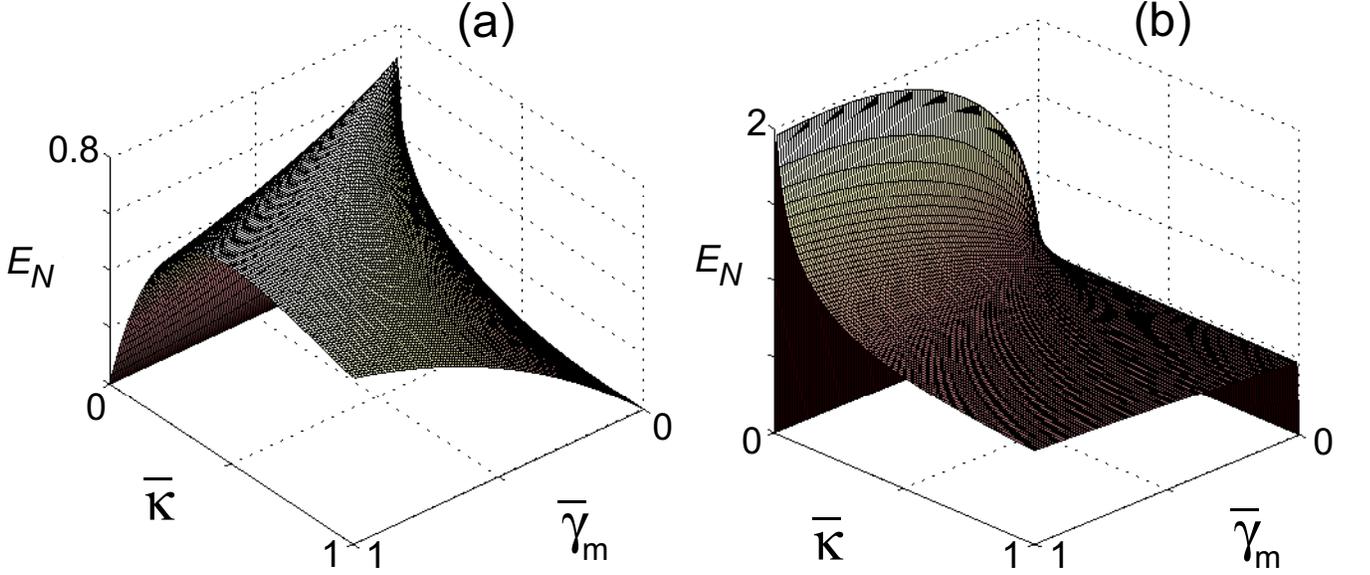}
\caption{Zero-temperature logarithmic negativity for (a) driving on the blue sideband, (b) driving on the red sideband. $\bar{\kappa}=\kappa
\times 10^{-6}{\rm s}^{-1}$, $\bar{\gamma}_m=\gamma_m \times 10^{-6}{\rm s}^{-1}$} \label{fig2}
\end{figure*}

\begin{figure*}[th!]
\resizebox{\textwidth}{!} {\includegraphics[scale=0.5]{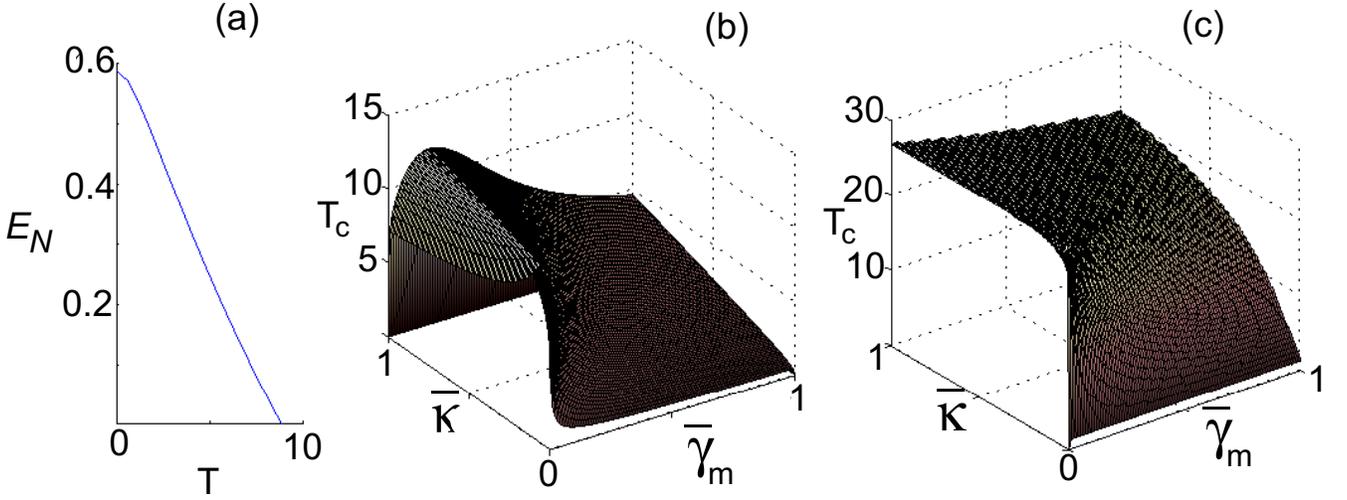}} \caption{(a) Typical temperature, $T$ (mK) dependence of the logarithmic
negativity (in this case, driving on the blue sideband with $\kappa = \gamma_m = 5\times 10^4 s^{-1}$). (b) $T_c$ (mK) as a function of damping
for driving on the blue sideband. (c) $T_c$  (mK)as a function of damping for driving on the red sideband. $\bar{\kappa}=\kappa \times
10^{-6}{\rm s}^{-1}$, $\bar{\gamma}_m=\gamma_m \times 10^{-6}{\rm s}^{-1}$ } \label{fig4}
\end{figure*}

\subsection{Results in the rotating wave approximation}

The results in the regime $\omega_m \gg G,\kappa$ may be understood with the aid of a rotating wave approximation (RWA) calculation. We will
refer to this as {\em weak driving} as $G$ depends on $n_s$, the intracavity photon number,  which itself depends on the strength of the driving
field, $E$.  It is then useful to introduce the nanomechanical annihilation operator $\delta b=(\delta q+i\delta p)/\sqrt{2}$, such that
Eqs.~(\ref{lle3a}) and (\ref{lle3b}) are equivalent to
\begin{equation}\label{bequat}
\delta \dot{b}=-i\omega_m \delta b - \frac{\gamma_m}{2} \left(\delta
b-\delta b^{\dagger}\right)+ i\frac{G}{2} \left( \delta a^{\dag}+
\delta a \right)+ \frac{\xi}{\sqrt{2}}
\end{equation}
and the whole system is described in terms of annihilation and creation fluctuation operators by Eqs.~(\ref{modelin}) and (\ref{bequat}). We now
move to a further interaction picture by introducing the slowly-moving tilded operators $\delta b(t)= \delta\tilde{b}(t)e^{-i\omega_m t}$ and
$\delta a(t)= \delta \tilde{a}(t)e^{-i \Delta t}$. They obey the QLEs
\begin{eqnarray}
\delta \dot{\tilde{b}}&=&- \frac{\gamma_m}{2} \left(\delta\tilde{b}-\delta \tilde{b}^{\dagger}e^{2i\omega_m t}\right)+
\frac{\xi e^{i\omega_m t}}{\sqrt{2}}, \nonumber \\
&& \ \ + i\frac{G}{2} \left( \delta\tilde{a}^{\dag}e^{i(\Delta+\omega_m) t}+\delta \tilde{a} e^{i(\omega_m-\Delta) t}\right) \label{mode2} \\
\delta \dot{\tilde{a}}&=&-\kappa\delta \tilde{a} +i \frac{G}{2}\left( \delta \tilde{b}^{\dag}e^{i(\Delta+\omega_m) t}+\delta\tilde{b}e^{i(\Delta-\omega_m) t}\right) \nonumber \\
&& \ \ +\sqrt{2\kappa} a^{in}e^{i \Delta t} .\label{modemech2}
\end{eqnarray}
The RWA allows us to ignore terms rotating at $\pm 2 \omega_m$ and use $\Delta \approx \Delta_0$. Then, for driving on the blue sideband we have
$\omega_c \simeq \omega_0-\omega_m$ and
\begin{eqnarray}
\delta \dot{\tilde{b}}&=&- \frac{\gamma_m}{2} \delta \tilde{b}+i\frac{G}{2}\delta \tilde{a}^{\dag} + \sqrt{\gamma_m} b^{in},  \label{mode3} \\
\delta \dot{\tilde{a}}&=&-\kappa\delta \tilde{a} +i \frac{G}{2}\delta \tilde{b}^{\dag} +\sqrt{2\kappa} \tilde{a}^{in}, \label{mode4}
\end{eqnarray}
and for driving on the red sideband we have $\omega_c \simeq \omega_0+\omega_m$ and
\begin{eqnarray}
\delta \dot{\tilde{b}}&=&- \frac{\gamma_m}{2} \delta \tilde{b}+i\frac{G}{2}\delta \tilde{a} + \sqrt{\gamma_m} b^{in},  \label{mode3prime} \\
\delta \dot{\tilde{a}}&=&-\kappa\delta \tilde{a} +i \frac{G}{2}\delta \tilde{b} +\sqrt{2\kappa} \tilde{a}^{in} . \label{mode4prime}
\end{eqnarray}
Note that $\tilde{a}^{in}(t)=a^{in}(t)e^{i \Delta t}$, possessing the same correlation function as $a^{in}(t)$, and
$b^{in}(t)=\xi(t)e^{i\omega_mt}/\sqrt{2}$ which, in the limit of large $\omega_m$, acquires the correlation functions \cite{gard}
\begin{eqnarray}
\langle b^{in,\dag}(t)b^{in}(t')\rangle &=& \bar{n}_b \delta (t-t'),\label{input1}\\
\langle b^{in}(t)b^{in,\dag}(t')\rangle &=& \left[\bar{n}_b+1\right]\delta (t-t'). \label{input2}
\end{eqnarray}
From Eqs.~(\ref{mode3})-(\ref{mode4}) we see that, for driving on the blue sideband, the cavity mode and nanomechanical mode play the role of
the signal and the idler of a nondegenerate parametric oscillator, characterized by an interaction term
$\delta\tilde{b}^{\dag}\delta\tilde{a}^{\dag}+ \delta\tilde{a}\delta\tilde{b}$. Therefore, it can generate entanglement. However, from
Eqs.~(\ref{mode3prime})-(\ref{mode4prime}), in the red sideband case the two modes are coupled by the beamsplitter-like interaction
$\delta\tilde{b}^{\dagger}\delta\tilde{a}+ \delta\tilde{a}^{\dagger}\delta\tilde{b}$, which is not able to entangle modes starting from
classical input states \cite{kim}.

Now introduce tilded quadrature operators $\delta \tilde{X}=(\delta \tilde{a}+\delta \tilde{a}^{\dag})/\sqrt{2}$ and $\delta \tilde{Y}=(\delta
\tilde{a}-\delta \tilde{a}^{\dag})/i\sqrt{2}$, with corresponding input noise operators $X^{in}=(\tilde{a}^{in}+\tilde{a}^{in,\dag})/\sqrt{2}$,
$Y^{in}=(\tilde{a}^{in}-\tilde{a}^{in,\dag})/i\sqrt{2}$, $q^{in}=(b^{in}+b^{in,\dag})/\sqrt{2}$, and $p^{in}=(b^{in}-b^{in,\dag})/i\sqrt{2}$. We
again obtain a system of the form (\ref{compact}), now with $u^T(t) = ( \delta \tilde{q}(t), \delta \tilde{p}(t)), \delta \tilde{X}(t), \delta
\tilde{Y}(t) $, $n^T(t)= (\sqrt{\gamma_m}q^{in}(t),\sqrt{\gamma_m}p^{in}(t), \sqrt{2\kappa}X^{in}(t), \sqrt{2\kappa}Y^{in}(t) )$ and
\begin{equation} \label{RWADynamics}
  A \equiv A^{\pm}=\frac{1}{2} \left(\begin{array}{cccc}
    -\gamma_m & 0 & 0 & \pm G \\
    0 & -\gamma_m & G & 0 \\
    0 & \pm G & -2\kappa & 0 \\
    G & 0 & 0 & -2\kappa
  \end{array}\right),
\end{equation}
where the upper (lower) sign corresponds to the blue (red) sideband case. For driving on the blue sideband, the stability condition of
Eq.~(\ref{stab1}) simplifies in the RWA limit to
\begin{equation} \label{StabRWA}
G < \sqrt{2\gamma_m\kappa },
\end{equation}
while the system is unconditionally stable for driving on the red sideband. For the symmetrically-ordered correlation matrix, we obtain an
equation of the form (\ref{lyaprwa}), which can be solved to give a matrix of the form
\begin{equation}\label{corremat2}
 V \equiv V^{\pm}=\left(\begin{array}{cccc}
    V^{\pm}_{11} & 0 & 0 & V^{\pm}_{14} \\
     0 & V^{\pm}_{11} & \pm V^{\pm}_{14} & 0 \\
    0 & \pm V^{\pm}_{14} & V^{\pm}_{33} & 0 \\
    V^{\pm}_{14} & 0 & 0 & V^{\pm}_{33}
  \end{array}\right),
\end{equation}
where
\begin{subequations}
\label{corrematelemrwa}
\begin{eqnarray}
V_{11}^{\pm}&=& \bar{n}_b+\frac{1}{2}+\frac{2G^2 \kappa \left[\left(\bar{n}_a+1/2\right)\pm\left(\bar{n}_b+1/2\right)\right]}
{\left(\gamma_m+2\kappa\right)\left(2\gamma_m \kappa \mp G^2\right)},\\
V_{33}^{\pm}&=& \bar{n}_a+\frac{1}{2}+\frac{G^2 \gamma_m \left[\left(\bar{n}_b+1/2\right)\pm\left(\bar{n}_a+1/2\right)\right]}
{\left(\gamma_m+2\kappa\right)\left(2\gamma_m \kappa \mp G^2\right)},\\
V_{14}^{\pm}&=& \frac{2G \gamma_m \kappa \left[\left(\bar{n}_b+1/2\right)\pm\left(\bar{n}_a+1/2\right)\right]}
{\left(\gamma_m+2\kappa\right)\left(2\gamma_m \kappa \mp G^2\right)}.
\end{eqnarray}
\end{subequations}
Now $\det V_{corr}^{\pm}= \mp (V_{14}^{\pm})^2$, which is non-negative in the red sideband case, a sufficient condition for the separability of
bipartite states \cite{simon}. Thus, in the red sideband case (RWA regime), the steady-state is not entangled.

We can quantify the entanglement by proceeding along the lines of (\ref{logneg}) and (\ref{blocks}). We may reproduce the entanglement of Figures 2 and 3 for the blue sideband case, but we see no entanglement for the red sideband case. This is because the RWA regime puts us at a coupling far below the instability threshold.
When the blue sideband steady-state correlation matrix is symmetric (that is, $2 \kappa = \gamma_m$ and $\bar{n}_a=\bar{n}_b=\bar{n}$) we find
\begin{equation}
E_{\mathcal{N}}=\max [0,\ln \frac{1+G/2\kappa}{1+2\bar{n}}]\label{logneg2}
\end{equation}
This and the stability condition of Eq.~(\ref{StabRWA}) imply that entanglement vanishes when $2\bar{n} > 1$, and that the logarithmic
negativity is bounded above as $E_{\mathcal{N}} < \ln 2$. Comparison with Figure 2(a) shows that this is actually an upper bound in all cases.

We shall now consider the experimental accessibility of the parameters described above. The coupling of Eq.~(\ref{coupling}) may be calculated
by assuming $\beta =0.002$, $d=100\ {\rm nm}$, $\omega_m\approx 10^8\ {\rm s}^{-1}, \  \omega_c\approx 6 \times 10^{10}\ {\rm s}^{-1}$,  and
$m=10^{-15}\ {\rm kg}$, giving $G_0 \approx 18\ {\rm s}^{-1}$. For these parameters the equivalent capacitance is $C\approx 0.1pF$ and an
equivalent inductance of $L\approx 3nH$.

For driving on the red sideband and the largest damping considered $(\gamma_m = \kappa = 10^5s^{-1})$, stability requires
$G<\sqrt{\omega_m^2+\kappa^2}\approx \omega_m$ so $G< 10^8\ {\rm s}^{-1}$. Maximal coupling, before loss of stability, then corresponds to
$\alpha_s = 10^7$, or a peak driving potential of $136\ {\rm mV}$, which would be feasible.

For driving on the blue sideband, the stability condition in Eq.~(\ref{bluestab}), with $\kappa=\gamma_m$ and $\omega_m \gg\kappa$ stability
requires $G<\sqrt{2}\kappa \approx 10^5 s^{-1}$.  Maximal coupling then corresponds to corresponding to $\alpha_s=10^3$, and a corresponding
maximum voltage of $38\ \mu{\rm V}$. It should also be noted that the very lowest damping rates depicted in Figures 2 and 3 would not be
achievable, due to the finite quality factors of the resonator and cavity.

\section{Conclusions}

We have shown a scheme able to entangle at the steady state a nanomechanical resonator with a microwave cavity mode of a driven superconducting
coplanar waveguide. The nanomechanical resonator is capacitively coupled with the central conductor of the waveguide and the steady state of the
system, in an appropriate parameter regime, is entangled up to temperatures of tens of milliKelvin. We have explained how this can be achieved
by presenting an approximate treatment based on a rotating wave approximation.

Let us briefly discuss how to detect the steady state entanglement. From the above equations, and especially the correlation matrix of the
steady state in the RWA limit, Eq.~(\ref{corremat2}), it is clear that the entanglement appears as a correlation between $\delta \tilde{q}(t)$
and $\delta\tilde{Y}(t)$, and also as a correlation between $\delta\tilde{p}(t)$ and $\delta\tilde{X}(t)$. A measurement of entanglement thus
requires that we measure these correlation functions. This is not an easy matter as it will require highly efficient measurements of both the
nanomechanical resonator displacement and the field amplitudes in the microwave cavity. Methods based on single electron transistors now enable
a displacement measurement at close to the Heisenberg limit \cite{schwab}. Unfortunately measurements of the weak voltages on the coplanar
cavity are not yet quantum limited due to the need to amplify the signals prior to detection. This is not a fundamental problem and a number of
efforts are underway to do quantum limited heterodyne detection of the cavity fields. It thus seems likely that a direct measurement of the
entanglement between a mesoscopic massive object and an electromagnetic field may be demonstrated using the approach of this paper. This would
provide a path to entangling many nanomechanical resonators via a common microwave cavity field.

\section{Acknowledgements}

This work was supported by the European Commission (program QAP) and by the Australian Research Council. We would like to acknowledge Keith
Schwab for helpful advice.

\end{document}